\documentclass[12pt]{article}
\usepackage{graphicx}
\usepackage{cite}
\usepackage{subcaption}
\usepackage{wrapfig}

\def\napoli{Institute of Particle and Nuclear Physics, Faculty of Mathematics and Physics, Charles University, Prague, Czech Republic}
\def\support{\footnote{On behalf of the H1 Collaboration}}

\def\Title#1{\begin{center} {\Large #1 } \end{center}}
\def\Author#1{\begin{center}{ \sc #1} \end{center}}
\def\Address#1{\begin{center}{ \it #1} \end{center}}

\newenvironment{Abstract}{\begin{quotation}  }{\end{quotation}}
\newenvironment{Presented}{\begin{quotation} \begin{center} 
             PRESENTED AT\end{center}\bigskip 
      \begin{center}\begin{large}}{\end{large}\end{center} \end{quotation}}

\def\beq{\begin{equation}}
\def\eeq#1{\label{#1}\end{equation}}
\def\eeqn{\end{equation}}

\def\beqa{\begin{eqnarray}}
\def\eeqa#1{\label{#1}\end{eqnarray}}
\def\eeqan{\end{eqnarray}}

\let\bar=\overbar

\def\Dslash{\not{\hbox{\kern-4pt $D$}}}
\def\dslash{\not{\hbox{\kern-2pt $\del$}}}

\def\msb{{\bar{\ssstyle M \kern -1pt S}}}

\begin{document}

\newcommand{\dstar}{\ensuremath{D^*}}
\newcommand{\pom}{{I\!\!P}}
\newcommand{\reg}{{I\!\!R}}

\begin{titlepage}
% \pubblock

\vfill
\Title{Measurement of $D^*$ Production in Diffractive Deep Inelastic Scattering at HERA}
\vfill
\Author{Karel \v Cern\' y\support}
\Address{\napoli}
\vfill
\begin{Abstract}
Measurements of $D^{*}(2010)$ meson production in diffractive deep inelastic scattering $(5<Q^{2}<100~{\rm GeV}^{2})$ are presented which are based on HERA data recorded at a centre-of-mass energy $\sqrt{s} = 319{\rm~GeV}$ with an integrated luminosity of $287$ pb$^{-1}$ collected by the H1 detector. The reaction $ep \rightarrow eXY$ is studied, where the system $X$, containing at least one $D^{*}(2010)$ meson, is separated from a leading low-mass proton dissociative system $Y$ by a large rapidity gap. The kinematics of $D^{*}$ candidates are reconstructed in the $D^{*}\rightarrow K \pi\pi$ decay channel. The measured cross sections compare favourably with next-to-leading order QCD predictions, where charm quarks are produced via boson-gluon fusion and they are independently fragmented to the $D^{*}$ mesons. The calculations rely on the collinear factorisation theorem and are based on diffractive parton densities previously obtained by H1 from fits to inclusive diffractive cross sections. The data are further used to determine the diffractive to inclusive $D^{*}$ production ratio in deep inelastic scattering.
\end{Abstract}
\vfill
\begin{Presented}
Presented at EDS Blois 2017, Prague, \\ Czech Republic, June 26-30, 2017
\end{Presented}
\vfill
\end{titlepage}
\def\thefootnote{\fnsymbol{footnote}}
\setcounter{footnote}{0}

\section{Introduction}
\begin{wrapfigure}{r}{0.35\textwidth}
  \includegraphics[width=1.0\linewidth]{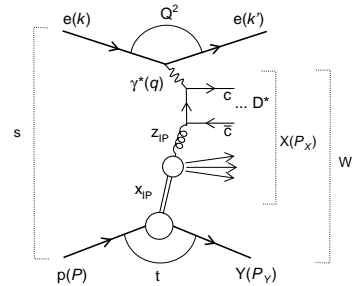}
  \caption{The leading order diagram for open charm production in DDIS at HERA.}% in the picture of collinear and proton vertex factorisation.}
  \label{fig:feyn}
\end{wrapfigure}

\par
The collinear factorisation~\cite{Collins:1997sr} approach is an effective tool to describe the diffractive deep inelastic scattering (DDIS) processes $ep \rightarrow eXY$, where the systems $X$ and $Y$ are separated by a large gap in rapidity due to the colourless exchange, often referred to as a pomeron ($\pom$).
 The diffractive parton distribution functions usually extracted from inclusive DDIS show that gluons constitute the main contribution to the DPDFs.
To date, analyses of HERA data support the validity of the collinear factorisation theorem in DDIS as evidenced by experimental results on inclusive production
, dijet production and $\dstar$ production. A brief account of a new measurement of $\dstar(2010)$ meson production in DDIS ~\cite{H1:2017bnb} is presented here. The $\dstar$ meson originates from the fragmentation of a charm quark, produced mainly via the boson-gluon-fusion ($\gamma^{*}g \rightarrow c\bar{c}$) at HERA energies, figure~\ref{fig:feyn}, and is therefore sensitive to the gluon content of the pomeron. The gluon content of the pomeron can be accessed directly and the collinear factorisation can be tested. Compared to the previous H1 publication~\cite{Aktas:2006up} the analysis presented corresponds to a sixfold increase in the integrated luminosity. 

\section{Kinematics of the deep inelastic scattering}
The standard DIS kinematics is described in terms of the well known invariants
\begin{eqnarray}
s = (k+P)^{2} ,\hspace*{0.7cm}
Q^2 = -q^{2}  ,\hspace*{0.7cm}
y = \frac{q\cdot P}{k\cdot P}  ,\hspace*{0.7cm}
W^2 = (q+P)^{2} ,\hspace*{0.7cm}
x = \frac{Q^{2}}{2\,q\cdot P} \ ,
\label{sqyx}
\end{eqnarray}
where the four-vectors are indicated in figure~\ref{fig:feyn}. The diffractive kinematics are defined as:
% The kinematic variables specific to diffraction are defined as follows:
\begin{eqnarray}
M_{X}^2 = (P_{X})^2 ,\hspace*{0.2cm}
M_{Y}^2 = (P_{Y})^2 ,\hspace*{0.2cm}
t = (P-P_{Y})^{2} ,\hspace*{0.2cm}
x_{\pom} = \frac{q\cdot (P-P_{Y})}{q\cdot P},\hspace*{0.2cm}
z_{\pom} = \frac{\hat{s} + Q^{2}} {M_{X}^{2} + Q^{2}},
\label{mxmytxpom}
\end{eqnarray}
where $M_{X}$ and $M_{Y}$ are the invariant masses of the systems $X$ and $Y$, respectively, $t$ is the squared four-momentum transfer at the proton vertex and $x_{\pom}$ the fraction of the proton's longitudinal momentum transferred to the system $X$ and $z_{\pom}$ represents, in leading order, the momentum fraction of $\pom$ participating in the $\gamma^{*}g \rightarrow c\bar{c}$ process, where $\hat{s}$ denotes its centre-of-mass energy squared.
 
\section{Monte Carlo Models and Fixed Order QCD Calculations}
\label{sec:MCandNLO}
The diffractive proton elastic $ep \rightarrow eX(\dstar)p$ and proton dissociative $ep \rightarrow eX(\dstar)Y$ are modelled with the RAPGAP Monte Carlo event generator~\cite{JUNG1995147}. Following a detailed H1 detector response simulation the samples are passed through the same analysis chain as used for data and are used to correct the data for detector effects.

Predictions for $\dstar$ cross sections in next-to-leading-order (NLO) QCD precision are obtained using HVQDIS~\cite{Harris:1995tu,Harris:1997zq}. The calculation relies on collinear factorisation using H1 2006 DPDF Fit B NLO parton density functions~\cite{Aktas:2006hy}. Massive charm quarks are produced via $\gamma^{*}$-gluon fusion. Fragmentation into $\dstar$ mesons is performed independently in the $\gamma^{*}p$ rest frame using the Kartvelishvili parameterisation with parameter values, as well as the uncertainties on them, suited for use with HVQDIS~\cite{Aaron:2008ac}. The factorisation and renormalisation scales are set to $\mu_{r} = \mu_{f} = \sqrt{Q^{2} + 4m_{c}^{2}}$ with the value $m_{c} = 1.5{\rm~GeV}$ for the charm pole mass. The variations by factors of $0.5$ and $2$ are used as the uncertainty of the scale's choice. The uncertainty introduced by the particular choice of $m_c$ is evaluated by varying it to  $1.3~{\rm GeV}$ and $1.7~{\rm GeV}$.

\section{Experimental technique}
A detailed description of the H1 detector can be in~\cite{Abt:1996hi}. The selection of DIS events is ensured by a scattered electron signal in the backward lead-scintillating fiber calorimeter. The decay products of the $\dstar$ meson are observed as tracks in the central tracker. Hadronic final state objects represent combined information from central tracker and LAr calorimeter.

The measured $Q^{2}$ and $y$ span $5 < Q^{2} < 100~{\rm GeV}^{2}$ and $0.02 < y < 0.65$ ranges, respectively. The large rapidity gap (LRG) selection of diffractive events is primarily provided by a cut on the position of the most forward cluster in the LAr calorimeter above the $800$ MeV energy threshold, $\eta_{\mbox{\tiny max}} < 3.2$, augmented by vetoes from forward detectors.
 The $x_{\pom} < 0.03$ range is used. The detection of $\dstar$ mesons is based on the full reconstruction of its decay products in the `golden channel': $D^{*+} \rightarrow D^{0} \pi^{+}_{slow} \rightarrow  (K^{-} \pi^{+}) \pi^{+}_{slow} + (C.C.)$ with a branching ratio of $\sim 2.7~\%$. The $K$ and $\pi$ candidate tracks are required to satisfy $p_{t} > 0.3 {\rm~GeV}$ transverse momentum cuts while the $\pi_{slow}$ track $p_{t} > 0.12~{\rm GeV}$ is required in the laboratory frame. An $80~{\rm MeV}$ mass window cut ensures consistency with $D^{0}$ hypothesis of $K$ and $\pi$ system. The $\dstar$ meson candidate kinematics is restricted to $p_{t,\dstar} > 1.5~{\rm GeV}$ and $\left| \eta_{\dstar} \right| < 1.5$. The variable $\Delta m = m(K^{\mp} \pi^{\pm} \pi^{\pm}_{slow})-m(K^{\mp} \pi^{\pm})$ is used to determine the $\dstar$ signal by means of simultaneous fits of the right and wrong charge combinations of the tracks for a better background shape determination yielding $N(D^{*}) = 1169 \pm 58$. The $N(\dstar)$ obtained from fits to the data in bins of event observables compare well with the simulation.%, see figure~\ref{fig:dmfit}.

%  % \begin{figure}[hhh]
% \begin{wrapfigure}{r}{0.5\textwidth}
% % \begin{subfigure}{.5\textwidth}
%  % \captionsetup{labelformat=empty}
%   \centering
%   \includegraphics[width=.98\linewidth]{bbb_control.eps}
%   % \caption{}
%   \label{fig:ctrlfig}
% % \end{subfigure}
% \caption{The $N(\dstar)$ from $\Delta m $ fits to data and simulation in bins of event observables.}
% \end{wrapfigure}
%  The $N(\dstar)$ obtained from fits to the data in bins of event observables are found to compare well with the simulation as can be seen in figure~\ref{fig:ctrlfig}, where fitted $N(\dstar)$ are shown differentially as a function of $Q^2$, $y$, ${\rm log}_{10}(x_{\pom})$, $z_{\pom}^{obs}$, $p_{t,\dstar}$ and $\eta_{\dstar}$.

% \begin{wrapfigure}{r}{0.5\textwidth}
%   \includegraphics[width=1.0\linewidth]{totaldatandstarrcwc.eps}
%   \caption{Result of $\Delta m $ fit to the total data statistics.}
%   \label{fig:dmfit}
% \end{wrapfigure}
% \begin{figure}[hhh]

\section{Results}
{The number of fitted $\dstar$ mesons is corrected for detector effects, branching ratio of the golden channel, contribution of other $\dstar$ decays and higher order QED processes at the lepton vertex. The phase space of the measurement reads $5 < Q^{2} < 100~{\rm GeV}^{2}$, $0.02 < y < 0.65$, $x_{\pom} < 0.03$, $p_{t,\dstar} > 1.5~{\rm GeV}$ and $\left| \eta_{\dstar} \right| < 1.5$. Since $M_{Y}$ and $t$ are not measured directly an extrapolation is performed
% \begin{wrapfigure}{r}{0.49\textwidth}
% \captionsetup{labelformat=empty}
%   \includegraphics[width=1.\linewidth]{bbb_bbbsinglediswithratio.eps}
%   % \caption{The leading order diagram for open charm production in DDIS at HERA in the picture of collinear and proton vertex factorisation.}
%   \caption{(a)}
%   \label{fig:xsec1}
% \end{wrapfigure}
to the range of $M_{Y} < 1.6~{\rm GeV}$ and $\left| t \right|~<~1~{\rm GeV}^{2}$. The integrated cross section of $\dstar$ production is measured to be $\sigma_{ep \rightarrow \,e Y \!X(\dstar) } = 314  \pm  23 {\rm (stat.) } \pm 35 {\rm (syst.) }~{\rm pb}$. The theoretical value calculated in next-to-leading order QCD reads $\sigma_{ep \rightarrow \,e Y \!X(\dstar)}^{\rm theory}$ $=265^{+54}_{-40}\,{\rm (scale)}$ $^{+68}_{-54}\,{\rm (m_{c})}$ $^{+7.0}_{-8.2} {\rm (frag.)}$ ${} ^{+31}_{-35}\,{\rm (DPDF)}~{\rm pb}$. Within  uncertainties both normalisation and shapes of the measured cross sections are reproduced by the theory, see figure~\ref{fig:xsec}. The experimental precision in the region of medium and highest $x_{\pom}$ and $z_{\pom}$, figure~\ref{fig:xsec2}, may provide a room further constraints on the gluon part of the DPDFs.
\begin{figure}[hhh]
% \captionsetup[subfigure]{labelformat=empty}
\begin{subfigure}{.5\textwidth}
  \centering
  \includegraphics[width=.99\linewidth]{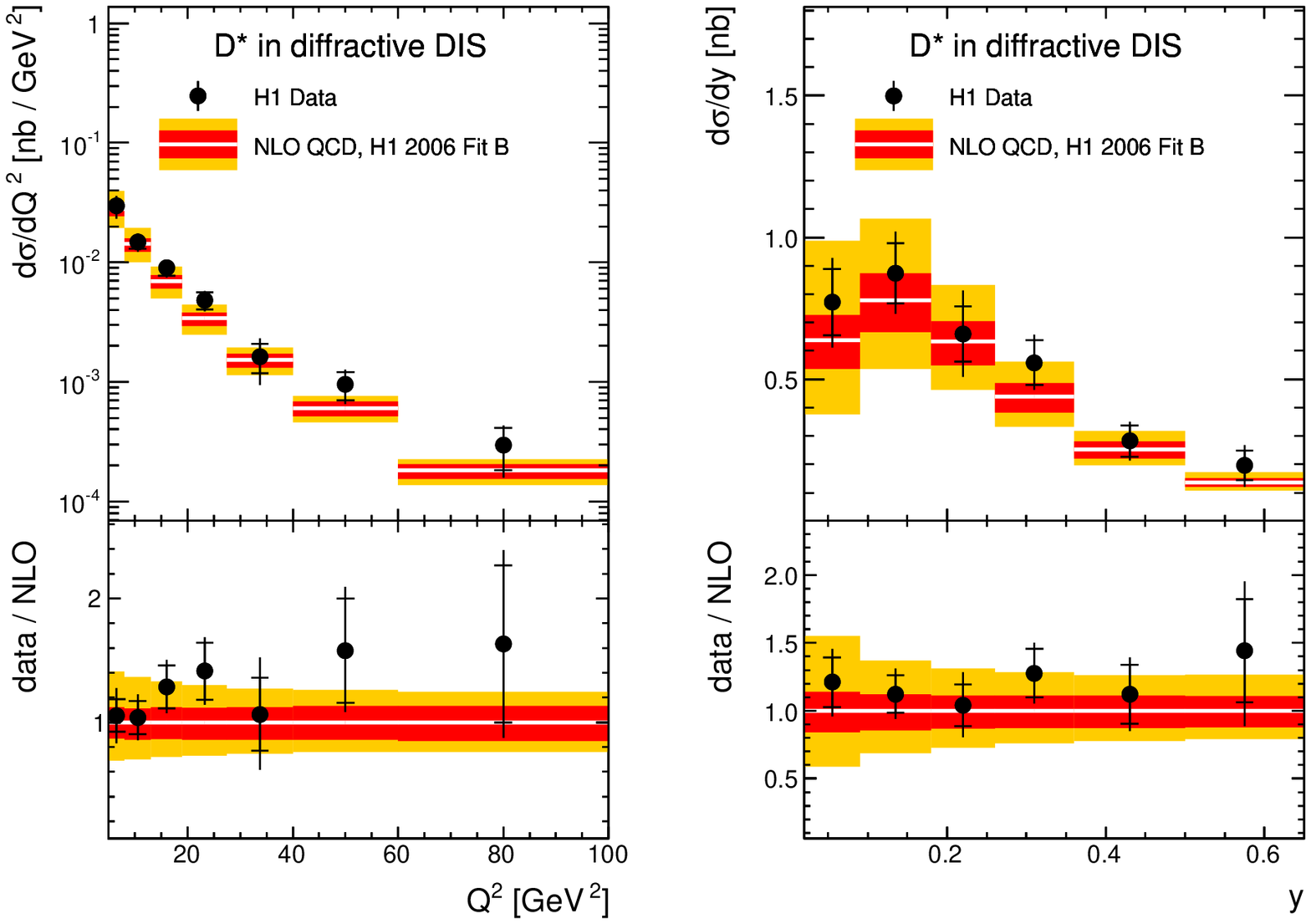}
  % \caption{(b)}
  \caption{}
  \label{fig:xsec1}
\end{subfigure}
\begin{subfigure}{.5\textwidth}
  \centering
  \includegraphics[width=.99\linewidth]{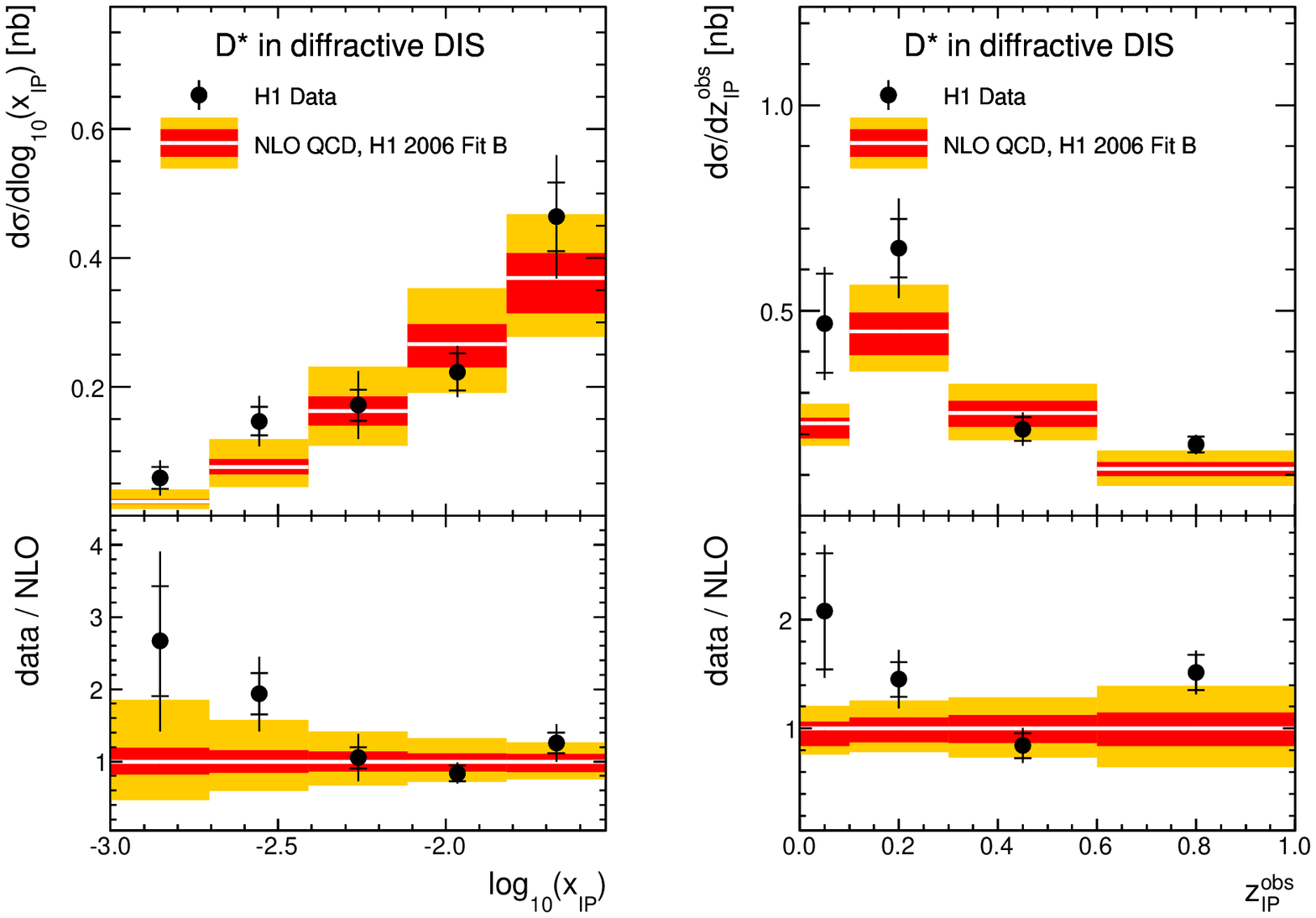}
  % \caption{(b)}
  \caption{}
  \label{fig:xsec2}
\end{subfigure}
% \begin{subfigure}{.5\textwidth}
%   \centering
%   \includegraphics[width=.99\linewidth]{bbb_bbbsingledstarwithratio.eps}
%   \caption{(c)}
%   \label{fig:xsec3}
% \end{subfigure}
\caption{Differential $\dstar$ production cross sections in DDIS and data to theory ratios as a function of; (a) $Q^2$ and $y$, (b) ${\rm log}_{10}(x_{\pom})$ and $z_{\pom}^{obs}$.}
\label{fig:xsec}
\end{figure} 

}

\begin{figure}[hhh]
% \captionsetup[subfigure]{labelformat=empty}
\begin{subfigure}{.45\textwidth}
  \centering
  \includegraphics[width=.99\linewidth]{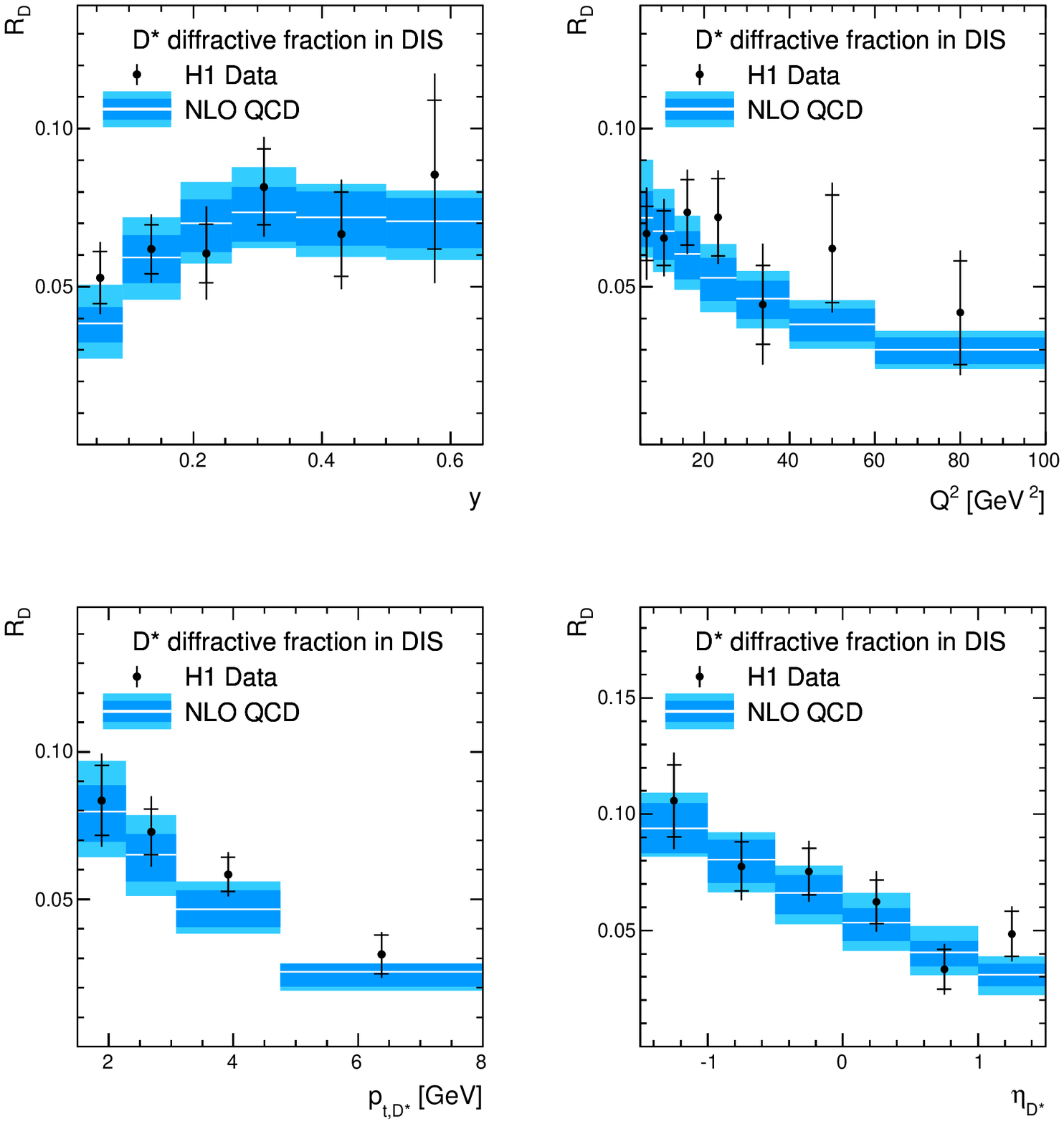}
  \caption{}
  \label{fig:rd1}
\end{subfigure}
\quad 
\begin{subfigure}{.5\textwidth}
  \centering
  \includegraphics[width=.99\linewidth]{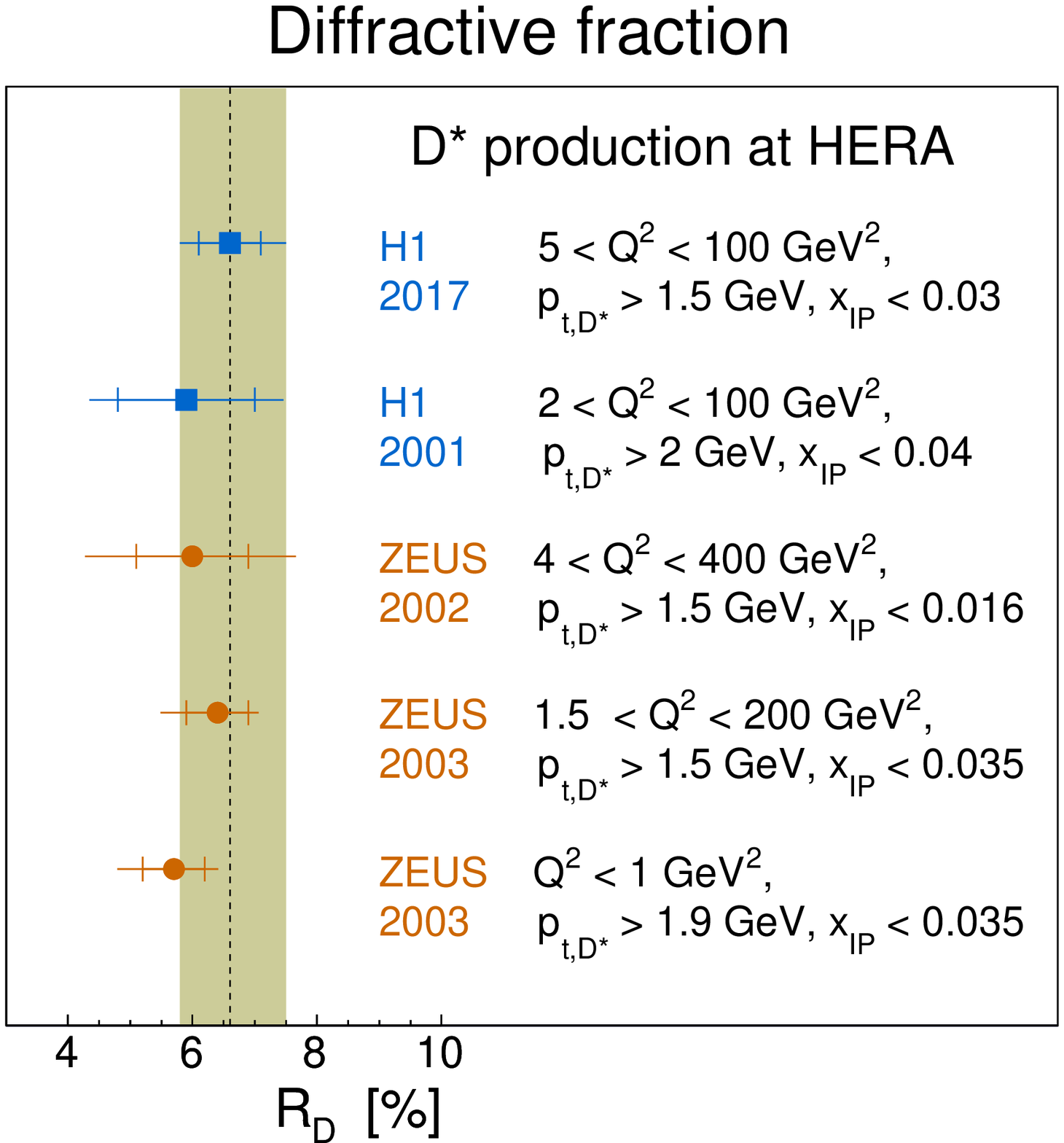}
  \caption{}
  \label{fig:rd2}
\end{subfigure}
\caption{The diffractive fraction $R_D$ as a function of $y$, $Q^2$, $p_{t,\dstar}$ and $\eta_{\dstar}$ (a). Integrated $R_D$ compared with previous measurements from HERA (b).}
\label{fig:rd}
\end{figure}

The fraction of diffractive~contribution~to the $\dstar$ production in DIS, $R_{D}$, is determined using also the H1 DIS results~\cite{Aaron:2011gp}. Integrated over the whole phase space the results read $R_{D} = 6.6 \pm 0.5 {\rm (stat)}~^{+0.9}_{-0.8} {\rm (syst)}\, \%$ and $R_{D}^{\rm\, theory} = 6.0 {}^{+1.0}_{-0.7} {\rm  (scale)}$ ${}^{+0.5}_{-0.4} {\rm (}m_c{\rm )}$ ${}^{+0.7}_{-0.8} {\rm  (DPDF)}$ ${}^{+0.02}_{-0.04} {\rm  (frag)} \,\%$ for the data and theoretical prediction, respectively, where the predictions profit from partial cancellation of the scale and $m_c$ uncertainties.
 % The uncertainties of the theoretical predictions are obtained from simultaneous variations of corresponding parameters in the DDIS and DIS regimes.
 The fractions measured differentially are shown in figure~\ref{fig:rd1}. 
Both the total and the differential data $R_D$ are consistent with theory showing strong kinematical dependence that can be explained limitations of the diffractive phase space domain. In figure~\ref{fig:rd2}, $R_D$ integrated over the full phase space is compared with previous measurements performed at HERA both in the DIS regime~\cite{Adloff:2001wr,Chekanov2002244,Chekanov20033} and in photoproduction~\cite{Chekanov2007}.

 % The average value of $R_{D}$ measured in this analysis is in agreement with the previous results and within the sizeable experimental uncertainties is observed to be largely independent of the varying phase space constraints in $x_{\pom}$, $Q^2$ and $p_{t,\dstar}$.

\section{Conclusions}
The integrated and differential cross sections of $\dstar(2010)$ production in diffractive deep inelastic scattering are measured. The measured cross sections are well described by theoretical predictions in next to leading order QCD. This supports the validity of collinear factorisation.

The measured diffractive fraction of $\dstar$ production cross section in deep inelastic scattering is in agreement with theoretical predictions in next-to-leading order QCD. The value of the diffractive fraction is found to depend on certain observables. It is, however, observed to be largely independent of other details of the phase space definition.

\bibliographystyle{JHEP}
\bibliography{mybibliography}{}

\providecommand{\href}[2]{#2}\begingroup\raggedright\begin{thebibliography}{10}

\bibitem{Collins:1997sr}
J.~C. Collins, \emph{{Proof of factorization for diffractive hard scattering}},
  \href{http://dx.doi.org/10.1103/PhysRevD.61.019902,
  10.1103/PhysRevD.57.3051}{\emph{Phys. Rev.} {\bfseries D57} (1998)
  3051--3056}, [\href{https://arxiv.org/abs/hep-ph/9709499}{{\ttfamily
  hep-ph/9709499}}].

\bibitem{H1:2017bnb}
{\scshape H1} collaboration, V.~Andreev et~al., \emph{{Measurement of $D^{*}$
  production in diffractive deep inelastic scattering at HERA}},
  \href{http://dx.doi.org/10.1140/epjc/s10052-017-4875-9}{\emph{Eur. Phys. J.}
  {\bfseries C77} (2017) 340},
  [\href{https://arxiv.org/abs/1703.09476}{{\ttfamily 1703.09476}}].

\bibitem{Aktas:2006up}
{\scshape H1} collaboration, A.~Aktas et~al., \emph{{Diffractive open charm
  production in deep-inelastic scattering and photoproduction at HERA}},
  \href{http://dx.doi.org/10.1140/epjc/s10052-006-0206-2}{\emph{Eur. Phys. J.}
  {\bfseries C50} (2007) 1--20},
  [\href{https://arxiv.org/abs/hep-ex/0610076}{{\ttfamily hep-ex/0610076}}].

\bibitem{JUNG1995147}
H.~Jung, \emph{{Hard diffractive scattering in high energy ep collisions and
  the Monte Carlo Generator RAPGAP}}, {\emph{Comp. Phys. Comm.} {\bfseries 86}
  (1995) 147 -- 161}.

\bibitem{Harris:1995tu}
B.~W. Harris and J.~Smith, \emph{{Heavy quark correlations in deep inelastic
  electroproduction}},
  \href{http://dx.doi.org/10.1016/0550-3213(95)00256-R}{\emph{Nucl. Phys.}
  {\bfseries B452} (1995) 109--160},
  [\href{https://arxiv.org/abs/hep-ph/9503484}{{\ttfamily hep-ph/9503484}}].

\bibitem{Harris:1997zq}
B.~W. Harris and J.~Smith, \emph{{Charm quark and $D^{*\pm}$ cross-sections in
  deeply inelastic scattering at HERA}},
  \href{http://dx.doi.org/10.1103/PhysRevD.57.2806}{\emph{Phys. Rev.}
  {\bfseries D57} (1998) 2806--2812},
  [\href{https://arxiv.org/abs/hep-ph/9706334}{{\ttfamily hep-ph/9706334}}].

\bibitem{Aktas:2006hy}
{\scshape H1} collaboration, A.~Aktas et~al., \emph{{Measurement and {QCD}
  analysis of the diffractive deep-inelastic scattering cross-section at
  HERA}}, \href{http://dx.doi.org/10.1140/epjc/s10052-006-0035-3}{\emph{Eur.
  Phys. J.} {\bfseries C48} (2006) 715--748},
  [\href{https://arxiv.org/abs/hep-ex/0606004}{{\ttfamily hep-ex/0606004}}].

\bibitem{Aaron:2008ac}
{\scshape H1} collaboration, F.~D. Aaron et~al., \emph{{Study of Charm
  Fragmentation into {$D^{*\pm}$} Mesons in Deep-Inelastic Scattering at
  HERA}}, \href{http://dx.doi.org/10.1140/epjc/s10052-008-0792-2}{\emph{Eur.
  Phys. J.} {\bfseries C59} (2009) 589--606},
  [\href{https://arxiv.org/abs/0808.1003}{{\ttfamily 0808.1003}}].

\bibitem{Abt:1996hi}
{\scshape H1} collaboration, I.~Abt et~al., \emph{{The H1 detector at HERA}},
  \href{http://dx.doi.org/10.1016/S0168-9002(96)00893-5}{\emph{Nucl. Instrum.
  Meth.} {\bfseries A386} (1997) 310--347, ibid. 348--396}.

\bibitem{Aaron:2011gp}
{\scshape H1} collaboration, F.~D. Aaron et~al., \emph{{Measurement of
  $D^{*\pm}$ Meson Production and Determination of $F_2^{c\bar{c}}$ at low
  $Q^2$ in Deep-Inelastic Scattering at HERA}},
  \href{http://dx.doi.org/10.1140/epjc/s10052-011-1769-0,
  10.1140/epjc/s10052-012-2252-2}{\emph{Eur. Phys. J.} {\bfseries C71} (2011)
  1769}, [\href{https://arxiv.org/abs/1106.1028}{{\ttfamily 1106.1028}}].

\bibitem{Adloff:2001wr}
{\scshape H1} collaboration, C.~Adloff et~al., \emph{{$D^{*\pm}$ meson
  production in deep inelastic diffractive interactions at HERA}},
  \href{http://dx.doi.org/10.1016/S0370-2693(01)01155-8}{\emph{Phys. Lett.}
  {\bfseries B520} (2001) 191--203},
  [\href{https://arxiv.org/abs/hep-ex/0108047}{{\ttfamily hep-ex/0108047}}].

\bibitem{Chekanov2002244}
{\scshape ZEUS} collaboration, S.~Chekanov et~al., \emph{Measurement of
  diffractive production of {$D^{*\pm}$} (2010) mesons in deep-inelastic
  scattering at {HERA}}, {\emph{Phys. Lett. B} {\bfseries 545} (2002) 244 --
  260}.

\bibitem{Chekanov20033}
{\scshape ZEUS} collaboration, S.~Chekanov et~al., \emph{Measurement of the
  open-charm contribution to the diffractive proton structure function},
  {\emph{Nucl. Phys. B} {\bfseries 672} (2003) 3--35}.

\bibitem{Chekanov2007}
{\scshape ZEUS} collaboration, S.~Chekanov et~al., \emph{{Diffractive
  photoproduction of $D^{*\pm}$ (2010) at HERA}}, {\emph{Eur. Phys. J.}
  {\bfseries C51} (2007) 301--315}.

\end{thebibliography}\endgroup

\end{document}